\documentclass[letterpaper,twocolumn,10pt]{article}
\usepackage{usenix-2020-09_v3}

\usepackage[english]{babel}
\usepackage{blindtext}
\usepackage{multirow}
\usepackage{amsmath,algorithm,algpseudocode}
\algnewcommand\Input{\item[\textbf{Input:}]}
\algnewcommand\Output{\item[\textbf{Output:}]}
\algnewcommand\Type{\item[\textbf{Type Definition:}]}
\usepackage{circledsteps}
\pgfkeys{/csteps/fill color=black}
\pgfkeys{/csteps/inner color=white}

\usepackage{graphicx}

\usepackage{amsmath}
\usepackage{amssymb}
\usepackage{pifont}

\newcommand{\cmark}{\ding{51}} % check mark
\newcommand{\xmark}{\ding{55}} % cross mark
\DeclareMathOperator*{\ubigvee}{\underline{\bigvee}}
\usepackage{hyperref}
\usepackage{outlines}
\usepackage{xcolor}
\usepackage{comment}
\usepackage{epigraph}
\usepackage{enumitem}
\usepackage{epigraph}
\usepackage{xspace}
\usepackage{soul}
\usepackage[normalem]{ulem}
\usepackage{booktabs}
\usepackage{natbib}

\usepackage{caption}
\usepackage{enumitem}
\usepackage[small,compact]{titlesec}

\usepackage{amsmath}
\usepackage{circledsteps}
\usepackage{colortbl}
\usepackage{enumitem}

\usepackage{xcolor}
\usepackage{listings}

\usepackage{algorithm}
\usepackage{algpseudocode}
\usepackage{graphicx}
\usepackage{textcomp}

\usepackage{subcaption}

\definecolor{archiecodegreen}{rgb}{0.24,0.48,0.48}
\definecolor{archiecodekeyword}{rgb}{0.00,0.50,0.00}
\definecolor{archiecodestring}{rgb}{0.73,0.13,0.13}
\definecolor{archiecodenumber}{rgb}{0.40,0.40,0.40}
\lstdefinestyle{archiecode}{
  basicstyle=\ttfamily\footnotesize,
  commentstyle=\itshape\color{archiecodegreen},
  keywordstyle=\bfseries\color{archiecodekeyword},
  stringstyle=\color{archiecodestring},
  numberstyle=\scriptsize\color{archiecodenumber},
  numbers=left,
  numbersep=5pt,
  breaklines=true,
  breakatwhitespace=false,
  showstringspaces=false,
  keepspaces=true,
  columns=fullflexible,
  upquote=true,
  tabsize=2,
  frame=none,
  xleftmargin=0pt,
  xrightmargin=0pt,
  aboveskip=4pt,
  belowskip=4pt
}
\floatstyle{plain}
\newfloat{listing}{tbp}{lol}
\floatname{listing}{Listing}
\newcommand{\setminted}[2][]{}
\newcommand{\usemintedstyle}[1]{}

\makeatletter
{\catcode`\!=8 % funny catcode so ! will be a delimiter
 \catcode`\Q=3 % funny catcode so Q will be a delimiter
\long\gdef\given#1{88\fi\Ifbl@nk#1QQQ\empty!}
\long\gdef\blank#1{88\fi\Ifbl@nk#1QQ..!}% if null or spaces
\long\gdef\nil#1{\IfN@Ught#1* {#1}!}% if null
\long\gdef\IfN@Ught#1 #2!{\blank{#2}}
\long\gdef\Ifbl@nk#1#2Q#3!{\ifx#3}% same as above
}
\makeatother
\def\expblank{\expandafter\blank\expandafter} % additional macro

\usepackage[textsize=tiny,textwidth=0.6in]{todonotes}

\newcommand{\allnotes}[1]{}
\newcommand{\noteakshay}[1]{\allnotes{\todo[color=cyan!50]{[Akshay: #1]}}}

\newcommand{\va}[1]{\allnotes{\todo[color=purple!50]{[Venkat: #1]}}}

\if\expblank{\allnotes{a}}%

\else

\fi

\newcommand{\sysname}{Kepler\xspace}

\hypersetup{pdfstartview=FitH,pdfpagelayout=SinglePage}

\newcommand{\cut}[1]{}
\newcommand{\paragrapha}[1]{\vspace{0.05in}\noindent{\bf #1.}}

\begin{document}
\title{Assistants, Not Architects: The Role of LLMs in Networked Systems  Design}

\author{
    {\rm Pratyush Sahu}\\
    Georgia Tech\\
    %\texttt{email@email.com}
    \and
    \and
    {\rm Rahul Bothra}\\
    UIUC\\
    %\texttt{email2@email.com}
    \and
    \and
    {\rm Venkat Arun}\\
    UT Austin\\
    \and
    \and
    {\rm Brighten Godfrey}\\
    UIUC\\
    \and
    {\rm Akshay Narayan}\\
    Brown University\\
    \and
    {\rm Ahmed Saeed}\\
    Georgia Tech\\
}

\maketitle

\begin{abstract}
\noteakshay{need to rewrite}

Designing the architecture of modern networked systems requires navigating a large, combinatorial space of hardware, systems, and configuration choices with complex cross-layer interactions. Architects must balance competing objectives such as performance, cost, and deployability while satisfying compatibility and resource constraints, often relying on scattered rules-of-thumb drawn from benchmarks, papers, documentation, and expert experience. This raises a natural question: can large language models (LLMs) reliably perform this kind of architectural reasoning? We find that they cannot. While LLMs produce plausible configurations, they frequently miss critical constraints, encode incorrect assumptions, and exhibit ``stickiness'' to familiar patterns. A natural workaround--iterative validation via simulation or experimentation--is often prohibitively expensive at scale and, in many cases, infeasible, particularly when comparing hardware-dependent alternatives.

Motivated by this gap, we present \sysname, a lightweight reasoning framework for architecture design that combines structured, expert-driven specifications with SMT-based optimization. \sysname encodes architecturally significant properties--requirements, incompatibilities, and qualitative trade-offs--about systems, hardware, and workloads as constraints, and synthesizes feasible designs that optimize user-defined objectives. It operates at an abstract level, capturing ``rules-of-thumb'' rather than detailed system behavior, enabling tractable reasoning while preserving key interactions, and provides explanations for its decisions. Through experiments and case studies, we show that \sysname uncovers interactions missed by LLMs and supports systematic, explainable design exploration.
\end{abstract}

\section{Introduction}\label{s:intro}

The modern infrastructure architect---spanning both network and application domains---faces a design problem that is simultaneously broader and more entangled than it was even a decade ago. Beyond choosing switches, NICs, and middleboxes, network architects now routinely decide among congestion-control and traffic-engineering regimes, in-network versus host-based processing, hardware offloads versus software datapaths, and a growing ecosystem of controllers and management planes. 
At the same time, application architects make configuration decisions that can be constrained by and impact network design: selecting a service-mesh data plane and control-plane topology; choosing load-balancing policies, retries, timeouts,
and mutual TLS; deciding between horizontal pod autoscaling, request-based scaling, queue-length scaling, or predictive approaches; and picking placement, sharding, caching, and deployment strategies. 
We refer to the resulting combination of hardware, systems, and configuration choices as an \emph{architecture}.

Network architecture design is difficult because of the combinatorial nature of the design space and the fact that choices rarely compose cleanly. Over the past three decades, our community has produced a rich set of systems---from end-to-end congestion control to orchestrators, traffic engineering solutions, in-network packet scheduling, load balancers, and accelerators---each targeting a subset of objectives. 
For example, choosing mTLS everywhere can simplify compliance and improve security posture, but may shift CPU budgets and influence whether kernel bypass or hardware offloads are necessary to meet throughput goals. 
Deploying a load balancer at the edge may simplify policy enforcement (e.g., colocating firewalling and termination), but may constrain routing, observability, and failure-domain isolation. 
Autoscaling policies can reduce cost under variable demand, but their effectiveness depends on workload burstiness, cold-start behavior, queueing dynamics, and the placement and topology of mesh sidecars. 
These are not corner cases; they are the common path. 
Architects thus must continually reason about \emph{interactions} across layers: application policy affects traffic; traffic affects congestion and load balancing; load balancing affects tail latency; tail latency affects the autoscaler; and the autoscaler changes traffic and placement again. 
Each of those components affects network design choices and configuration. 

The community has traditionally addressed these challenges by evaluating components in isolation: benchmarking implementations, simulating protocols, or analyzing models. 
While valuable, 
considering individual components doesn't
directly answer the architect’s central question: \emph{how do we put the pieces together?} 
Rather, answering this question requires understanding dozens of pairwise system interactions, represented eventually as the architect's experience and reasoning.
Instead, consider a hypothetical oracle that could capture the essential interaction rules across layers and components. 
An architect could ask whether a candidate architecture satisfies constraints (hardware compatibility, policy requirements, operational limits), whether it meets performance and cost objectives, and what minimal changes would restore feasibility when constraints change. 
Different teams could use the oracle to ensure cross-compatibility of independently developed subsystems (e.g., mesh policy, transport choices, load balancer placement, and NIC capabilities). 
Developers could use the oracle to understand which deployment contexts their systems are well suited for, and which constraints routinely make them inapplicable.

While building a complete oracle would be difficult, 
recent advances in large language models (LLMs) appear to offer a shortcut. 
LLMs can summarize design documents, enumerate plausible alternatives, and generate architecture diagrams that read like expert recommendations. 
Indeed, recent work has argued that systems problems are well suited to \emph{AI-driven} design loops that combine LLMs and expert verifiers, using simulations or experimentation, to produce complex system designs~\cite{cheng2025let, cheng2026ai, hamadanian2025glia}.
Yet, architecture design in production settings is neither a ``best-practices retrieval'' problem nor a search for plausible suggestions. In practice, it requires 
(i) \emph{optimization} within a trade-off space of architect-defined objectives (e.g., performance, cost, and maintainability),
(ii) \emph{nuanced reasoning} about interactions and second-order effects, 
(iii) \emph{explainability} so engineers can audit why a configuration is recommended, 
and (iv) \emph{transparency} about assumptions, constraints, and tradeoffs. 
A suggestion that is plausible in isolation is not actionable unless it 
is compatible with the rest of the deployment, 
satisfies hard constraints (e.g., hardware inventory, compliance, tail-latency SLOs), 
and is near-optimal with respect to stated objectives.

The findings we report in this paper suggest that while LLMs are valuable aids for information retrieval and design space exploration, they cannot not reliably architect common networked systems. 
In particular, we find that although LLMs can accurately produce specifications in well-structured domains such as datacenter hardware, they struggle to produce comprehensive and correct specifications for software systems, where subtle constraints, compatibility caveats, and conditional behaviors are critical. 
Their outputs frequently omit important properties and, more concerningly, include incorrect assertions that materially alter architectural decisions. 
This indicates that treating LLMs as end-to-end design agents is risky: the resulting specifications require careful human auditing, undermining the benefits of automation.

Motivated by this gap, this paper takes a preliminary and exploratory step toward building a network design oracle that combines LLMs' strength in interpreting natural-language descriptions with a \emph{lightweight reasoning engine}, \sysname, designed to make architectural reasoning nuanced, explainable, and transparent. 
Our key insight is that effective reasoning about architectural design questions requires representations that are \emph{structured}, yet deliberately \emph{shallow}. 
Fully modeling the detailed behavior of even one component is difficult; modeling detailed interactions among many components is intractable without abstraction. 
\sysname instead encodes (using SMT formulas) known facts---rules-of-thumb, requirements, incompatibilities, and conditional recommendations---about deployable systems, hardware components, and application workloads, without encoding implementation internals. 
\sysname thus serves as a machine- and human-readable compendium of the community's collective knowledge about systems. 
Machine-readability is essential 
to constrain and fortify LLM-based assistance. 
LLMs can extract relevant constraints and caveats from unstructured sources such as documentation and post-mortems. 
Then, a human expert can audit and improve those specifications during an encoding phase.
Later, \sysname can surface relevant facts during the design phase, accompanied by traceable explanations grounded in explicit rules rather than opaque model behavior. This paper builds on our prior work~\cite{our_hotnets} by fully specifying the reasoning system, refining its abstractions and interface, and providing a comprehensive evaluation of LLMs in architecture design.

This paper makes the following contributions:
\begin{enumerate}[noitemsep,topsep=0pt,leftmargin=*]
\item An evaluation of LLMs' suitability for network architecture design tasks. When used alone, LLMs have a $43\%$ failure rate when encoding cloud systems' properties (\S\ref{s:llm-specs-eval}) and, when producing designs, make clear mistakes leading to suboptimal latency (\S\ref{sec:case_study_1}), but with \sysname can help produce clear explanations for design choices.
\item \sysname, a lightweight reasoning engine that leverages LLMs for the tasks they are suited to: human-verified encoding and explaining structured output.
\item Two case studies spanning over $50$ system encodings using \sysname to explore architecture design decisions, revealing subtle cross-component interactions (``gotchas'') LLM-only baselines miss.
\end{enumerate}

\section{Architecture as an Optimization Problem}

Designing networked systems is fundamentally an optimization problem. System architects must navigate a large design space to balance competing objectives such as performance (e.g., throughput and latency), cost, scalability, maintainability, and operational complexity, subject to constraints imposed by hardware, software ecosystems, and organizational practices. Unlike classical optimization problems defined over precise objective functions, architectural optimization is qualitative and comparative: architects make decisions based on high-level metrics, abstractions informed by broad system knowledge the architect’s prior experience, risk tolerance, and the human-level strengths of their team.
Architectural decisions are inherently cross-layer. Application-level policies shape traffic patterns; traffic interacts with congestion control, load balancing, and scheduling mechanisms; these interactions ultimately influence tail latency, autoscaling behavior, and cost. As a result, evaluating a single design choice requires reasoning about multiple components and constraints simultaneously. The design space is combinatorial, and dependencies are often indirect, making exhaustive reasoning impractical.

In practice, architects cope with this complexity by reasoning at an abstract level. Rather than modeling systems in full operational detail, they focus on a small number of architecturally significant properties that rule out large portions of the design space. For example, when choosing a datacenter congestion control algorithm, an architect may focus on whether one option requires INT-enabled switches while another relies on dedicated QoS levels. Such properties are naturally expressed as simple predicates, while relative performance is often captured as an ordering informed by workload characteristics and prior experience rather than precise quantitative models. 
Quantitative reasoning is typically limited to metrics that are easy to specify and compare—such as cost in dollars, resource capacities, or simple piecewise-linear constraints—while quantities like tail latency are treated qualitatively due to the difficulty of predicting them without simulation or deployment.

This style of reasoning deliberately excludes low-level details that do not materially affect architectural decisions. From an architect’s standpoint, a small set of salient facts can suffice to check many properties at once, improving legibility and tractability while still capturing the core tradeoffs. Importantly, these abstractions also reflect the reality that many inputs—such as cost or operational complexity—are subjective and vary across organizations.

While system architecture design has traditionally been the exclusive domain of human experts, the combinatorial nature of the design space---with a growing number of roles and systems that can play each role---creates the need to track an unreasonable number of facts about each system as well as their interactions across layers, making manual exploration difficult and error-prone. 
These limitations motivate an explicit formulation of architecture design as an optimization problem over high-level system properties, qualitative preferences, and tractable constraints. 
One of our contributions is to articulate such a formulation clearly. By making the structure of architectural reasoning explicit, we provide a concrete basis for assessing whether LLMs can reliably reason about system architecture—or whether their strengths are limited to generating plausible but shallow designs.

\section{The Role of LLMs in Architecture Design}\label{ss:LLM-as-Architects}

To cope with the growing complexity of the design space, we highlight two complementary human roles.
The \emph{architect} possesses broad, cross-cutting knowledge of multiple components and their interactions, but may lack detailed or up-to-date understanding of each subsystem. 
In contrast, a \emph{subject-matter expert} has deep knowledge of a specific component or domain, including subtle constraints, operational caveats, and implementation details.
Effective architecture design requires combining these perspectives, as modern systems exhibit complex interactions that exceed the working memory of any single individual.

At first glance, architecting networked systems appears well aligned with LLMs' strengths. 
The task typically involves extracting system-level constraints and capabilities from heterogeneous, unstructured sources (e.g., documentation, design reports, and prior implementations) and reasoning about trade-offs to produce a coherent design. 
However, the details are important; we find that how exactly architecture design involves LLMs affects the outputs significantly.
We thus describe four points in this design space: 
(i) one-shot design,
(ii) iterative reasoning,
(iii) specification extraction,
and (iv) expert-driven specification (which \sysname uses).

\subsection{Design Space}

\paragrapha{One-shot architecture design} 
The simplest approach is for the architect to prompt an LLM to generate a complete architecture in a single pass, relying on the LLM to replace all experts. 
While convenient, past empirical studies already show that zero-shot or one-shot settings struggle with complex reasoning tasks, particularly those requiring multi-step or compositional reasoning or optimization~\cite{shojaee2025illusion}.
As a result, one-shot generation is unsuitable for production-quality architecture design.
We show examples of such reasoning problems in Section~\ref{sec:case_study_1}.

\paragrapha{Iterative reasoning}
A potential way to improve on one-shot design is naturally multi-shot design, or \emph{iterative reasoning}, where the model generates multiple candidate designs (via sampling), and subsequently evaluates or validates them. 
This paradigm includes techniques such as chain-of-thought reasoning, self-consistency sampling, and generate-verify loops, where an agentic application explores and filters multiple reasoning traces based on correctness criteria. 
While this approach improves robustness compared to one-shot prompting, it introduces significant practical challenges in the context of architecture design. 
Meaningful validation often requires \emph{heavyweight evaluation infrastructure}, such as detailed simulation environments, emulation frameworks, or even real-world experimentation. 
When hardware is involved, validation becomes even more challenging. 
Many architectural decisions depend on hardware-specific features (e.g., NIC capabilities, programmable switches, offloads), which the agentic application cannot faithfully evaluate without access to the actual hardware (or highly accurate models of it). 
Constructing such validation setups is often prohibitively expensive, time-consuming, or outright infeasible, especially when exploring many candidate designs.

\paragrapha{Specification extraction and formal optimization}
A more conservative approach instead uses LLMs only to translate unstructured inputs (e.g., documentation, design goals) into structured, machine-readable specifications, then uses a formal optimizer such as an SMT solver to process these specifications. 
This aligns with LLM-to-symbolic pipelines that combine natural language processing with tool calls to formal reasoning systems. 
For example, recent work has proposed using LLMs as intermediaries that translate unstructured descriptions into structured representations, which downstream analyzers or optimization engines can apply to domains such as robotics~\cite{haoplanning} or deductive reasoning problems~\cite{pan2023logic}.

This approach involves both (i) extracting accurate specifications of systems components and (ii) synthesizing appropriate abstractions to use with the formal optimizer (i.e., what details are important to encode?).
We evaluate the performance of a frontier LLM service, Gemini 3 Pro, on this task in \S\ref{s:llm-specs-eval} and find that the output is often inaccurate.

\paragrapha{Expert-driven specification with LLM assistance (\sysname)} 
We argue that the specification extraction design is more effective than more generic agentic application designs, but that its missing pieces are (i) a \emph{fixed} set of abstractions for use with an SMT-based optimizer (ii) expert supervision over LLM-produced system and workload specifications using this set of abstractions.

We thus design \sysname's set of abstractions to elide 
low-level details about systems' operation, which don't matter from an architectural standpoint.
Instead, we focus on modeling systems' high-level properties, which improves legibility, tractability, extensibility, and explainability. We refer to this level of specification as encoding ``rules-of-thumb''. 
Reasoning over these lightweight encodings is possible with finitely many variables, and users and LLMs can express queries as an existentially quantified formulas, i.e, ``Does there exist a choice of systems such that the following properties and constraints on cost, deployability, etc. are met?''. 
This is a decidable question that a SAT/SMT solver dealing with boolean and first-order predicates can answer. We choose to use Z3~\cite{z3}, a Satisfiability Modulo Theorem (SMT) Solver, as the reasoning framework over the rules-of-thumb that can efficiently describe, check, and optimize our models.

This approach allows users to leverage LLMs for tasks to which they are well suited:
(i) unstructured information retrieval, for example quantities from hardware specsheets; and
(ii) synthesizing SMT formulas to help answer users' natural-language queries about their system's design.
Using \sysname,
subject-matter experts (potentially themselves aided by LLMs) can specify and publish the properties of systems and hardware for reasoning and optimization.
Then, an architect can describe workloads they need to run and the hardware they have available.
\sysname can then interact with both these inputs to provide design recommendations backed by encoded facts and solver-provided reasoning.

\begin{table}[t]
\footnotesize
\centering
\begin{tabular}{p{2cm} c c c c}
\toprule
\textbf{System} & \textbf{Agree} & \textbf{Wrong} & \textbf{Additional} & \textbf{Missing} \\
\midrule
Docker                    & 2  & 0 & 0 & 0 \\
Containerd                & 2  & 3 & 0 & 1 \\
CRI-O                     & 0  & 4 & 2 & 0 \\
K8 Orchestrator    & 0  & 1 & 2 & 4 \\
Istio              & 1  & 3 & 1 & 4 \\
K8 Autoscaler                     & 1  & 0 & 2 & 3 \\
Linkerd     & 1  & 3 & 2 & 5 \\
KEDA                   & 0  & 2 & 2 & 2 \\
gRPC                   & 1  & 1 & 1 & 1 \\
Thrift                   & 0  & 3 & 2 & 2 \\
Docker Swarm                   & 3  & 0 & 0 & 0 \\
\midrule
\textbf{Total}            & \textbf{11 (\cmark)} & \textbf{20 (\xmark)} & \textbf{14 (+)} & \textbf{22 ($-$)} \\
\bottomrule
\end{tabular}
\caption{Summary of evaluation outcomes.}
\label{t:system_eval_summary}
\vspace{-0.1in}
\end{table}

\subsection{Evaluating LLM Specifications}\label{s:llm-specs-eval}
To assess the role LLMs should serve in system architecture design, we evaluate their ability to produce a comprehensive and correct system specification from a structured prompt. 
By specification, we mean a machine-readable description of the performance and operational objectives a system satisfies, together with the constraints and requirements on its deployment.
Our prompts describe high-level goals, operational constraints, and environmental assumptions and ask the model to enumerate the relevant architectural components, their properties, and the constraints governing their interaction. 
To best approximate the results a user today might see, we use the Gemini~3 Pro service to generate specifications and compare its output against specifications produced by a human expert. 
Our prompts require the LLM to provide verifiable references to support any statement its output makes about a system.
We consider two classes of systems: software systems for managing microservices-based applications, and datacenter hardware systems. We evaluate across 11 applications and 5 hardware systems. We select the applications to reflect well-known components commonly found in moderate- to large-scale microservice deployments. We select the hardware systems to span diverse device types, including compute nodes available in public cloud testbeds such as CloudLab \cite{duplyakin2019design}, GPU accelerators, and publicly available switch and router models.

Our results reveal a sharp contrast between these two domains. For datacenter hardware, the LLM service's output reproduces the expert's specification almost perfectly. In contrast, when generating specifications for microservices management software, the LLM's output diverges substantially from the expert baseline. 
To analyze this divergence, we compare individual properties appearing in the LLM-produced and human-produced specifications. Table~\ref{t:system_eval_summary} summarizes the results for the microservices-based applications.\va{I thought this table would have had some hardware examples too?}

We classify each property the LLM service produces into one of four categories: 
(1) \emph{Agree}, where the LLM and human expert identify the same property; 
(2) \emph{Wrong}, where the LLM produces an incorrect property; 
(3) \emph{Additional}, where the LLM identifies a correct property the human expert omitted; 
and (4) \emph{Missing}, where the LLM fails to identify a property the expert specified. We see that the LLM has a 
$43\%$
failure rate of all the specifications generated for systems. On the other hand, it achieves 
$100\%$ 
accuracy for hardware specifications. On average, the LLM produces about 1 additional specification that can be correctly validated, but it also misses, on average, 2 constraints that an expert would encode based on their knowledge.

Our results highlight that LLMs do not reliably produce of comprehensive and accurate specifications for complex software systems, especially in the presence of subtleties, caveats, and cross-cutting constraints.
These mistakes directly affect key architectural decisions.
Specifically, when we examined the references for incorrect specification components,
we found that the reference was a poor source or included only partial information.
For example, while the HPA autoscaler does require a container orchestrator, the LLM's output only allowed Kubernetes and excluded the compatible alternative Knative, despite its mention in the prompt.
Similarly, the LLM's encoding specifies that Linkerd and Docker cannot be used together; this misunderstanding was based on an obscure conflict between two specific versions of each system.

Overall, LLMs face a challenge when attempting to provide comprehensive system specifications due to the nature of online information about software systems, which is often fragmented across blogs, forums, and social media posts that may be incomplete, contradictory, or outdated. 
As a result, LLM-generated specifications are not only incomplete but can include confidently stated errors that require careful human auditing. This auditing can be time-consuming and error-prone, diminishing the practical benefit of full automation. 
We therefore conclude that while LLMs can be valuable tools to support human architects, we cannot rely on their output as the sole source of truth when designing complex systems. 
More broadly, our findings underscore the importance of maintaining system specifications in structured, machine-readable formats that support validation, auditing, and formal automated reasoning.

\begin{figure}[t]
\centering
\includegraphics[width=0.5\textwidth]{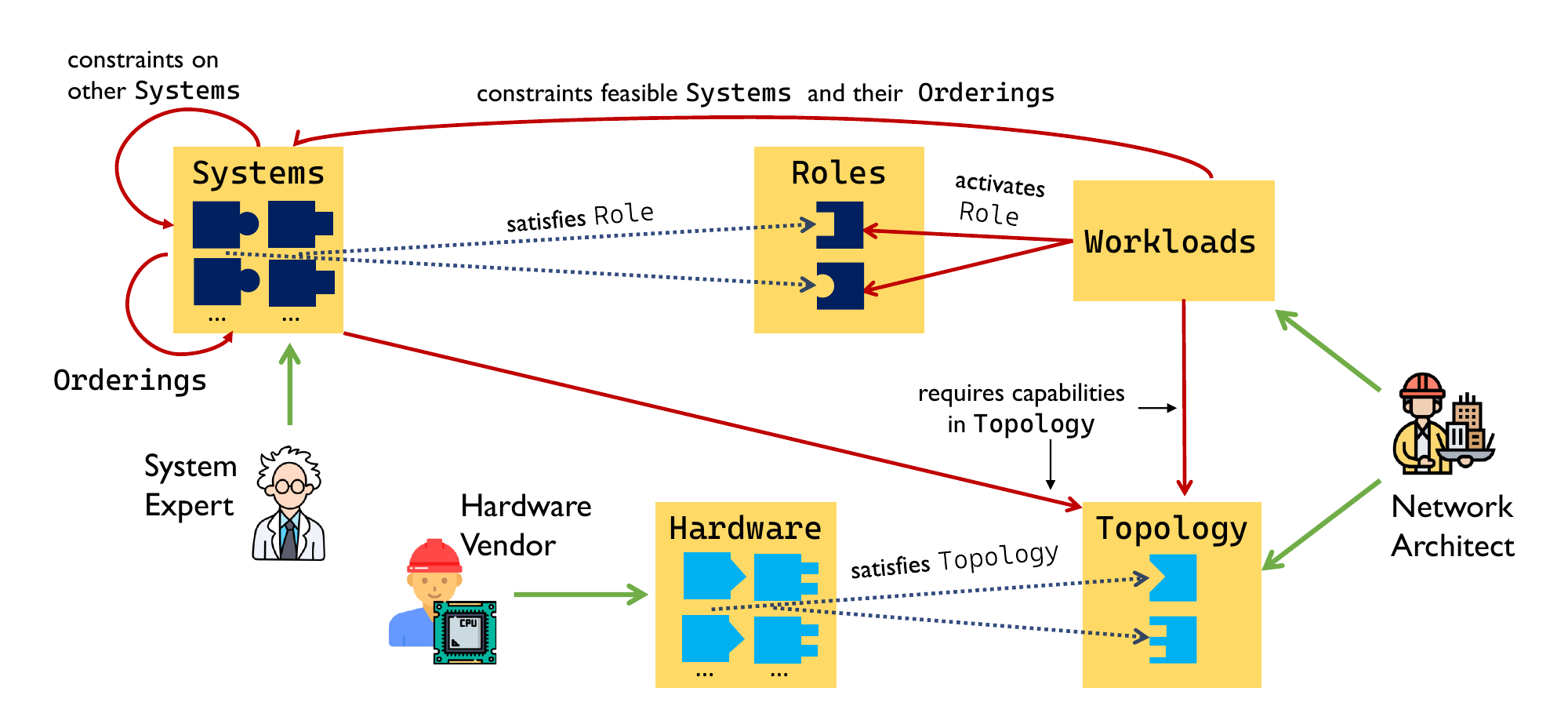}
\caption{\sysname's abstractions, which humans and LLMs can use to express system designs.}
\label{fig:arch}
\vspace{-0.2in}
\end{figure}

\section{\sysname Design}\label{s:design}

\sysname provides an interface in which system experts and architects communicate technical specifications and requirements of various kinds of systems and hardware. This section describes how the two groups denote their expertise over a formally defined interface understood by our reasoning framework.
This section focuses on the abstractions \sysname provides to guide reasoning about system designs; we discuss how LLMs improve its user experience for both system experts and hardware in \S\ref{s:casestudies}.

\subsection{\sysname Abstractions Overview}

Figure~\ref{fig:arch} summarizes the entities users can encode in \sysname. 
Given an architect’s 
workload (i.e., application characteristics and their hardware or software requirements), 
objectives, 
hardware acquisition costs, 
and existing inventory, 
\sysname synthesizes a design by selecting a 
system for each role and a 
hardware instance for each device
in the topology.

We designed these entities to apply across multiple layers of the protocol stack.
A device could encode hardware at a low level (e.g., ``switch'') or at a high level (e.g., ``VM'').
%(e.g., ``switch'' or ``NIC''). 
As a result, devices are simply a collection of boolean flags and numeric quantities denoting their capabilities (e.g., ``P4-programmable pipeline?'' or ``Pre-emptible compute?'') and properties (e.g., ``$16$ QoS levels'' or ``$8$ vCPUs'').
An individual hardware instance fulfills (or not) a device's contract (e.g., ``Tofino 2 switch'' or ``Amazon AWS EC2 \texttt{c8g.2xlarge}'').
Importantly, \sysname treats the quoted identifiers as symbolic labels: it does not encode any domain-specific semantics beyond the boolean flags and numeric attributes. 
For example, \sysname does not ``know'' what a switch is, only that workloads and systems require devices of type ``switch'' with some flags set (e.g., an encoding of DCTCP~\cite{dctcp} would add a requirement for ECN support). 
This abstraction is what we refer to as \emph{lightweight verification}.

Analogously, a role represents a functional slot that a system fills. 
A single system can fill multiple roles. 
Unlike devices, whose presence is implied, roles are not always active; instead, workloads require the presence of one or more roles.
For example, a workload could require a ``encryption in-flight'' role, which a service mesh system such as ``Cilium'' would fulfill.
Roles can imply other roles; e.g., ``encryption in-flight'' might in turn require a ``network stack'' (indeed, most \sysname designs will require a system such as ``Linux'' or ``Snap''~\cite{snap} that fulfills the ``network stack'' role).
Individual systems and workloads can impose constraints on devices such as those described above.

The modular decoupling of role and device from system and hardware
allows systems to query properties about each other to express complex logical dependencies. 
For instance, \sysname can encode the concept that a system like LEDBAT~\cite{LEDBAT}, satisfying a ``scavenger congestion control'' role can co-exist with a ``congestion control'' system like Cubic~\cite{cubic} or BBR~\cite{bbr} on the same QoS level only if the buffer size on the switch is large enough. 
Further, it cannot co-exist at all if the ``congestion control'' chosen by \sysname is Vegas~\cite{vegas} or Copa~\cite{copa}.

\paragrapha{Qualitative Partial Orderings}
\sysname's Ordering objects encode preferences between different systems with respect to a specific objective.  
A user defines an objective for a single role along a single axis (e.g., ``datacenter congestion control latency'' or ``key-value-store tail latency''). 
\sysname does not attempt to answer or encode quantitative questions (with the minor exception described above regarding finite resource slots);
instead, it encodes only \emph{partial orderings} over systems. 
These orderings may depend on workload characteristics, hardware capabilities, and interactions with other systems. For example, the relative performance of congestion control algorithms depends on whether a workload creates network traffic with incast.

Orderings provide a mechanism for subject-matter experts to encode rankings derived from quantitative evaluations, while still allowing architects to override these rankings based on their own preferences and context. Because users define orderings qualitatively, they can naturally capture attributes that are difficult to quantify, such as the extent to which a system requires application modifications or its level of software maturity. This allows distinguishing, for example, production-hardened systems from research prototypes in the design process.

Since \sysname must reconcile many orderings, it composes them using a lexicographic optimization scheme rather than numeric weighting, which would lack consistent real-world interpretation. 
Architects specify a priority ordering over objectives (see~\S\ref{s:interface}), such as cost, application throughput, network latency, and software maturity. 
This enables exploration of different trade-offs and expressing constraints such as ``latency must be at least as good as DCTCP.'' 
Further, partial orderings reflect user intent, so an LLM front-end for \sysname can write queries that explain why higher-ranked options for a particular ordering are not in the output.
This in turn can provide an opportunity for the user to revisit their orderings or constraints.
Generally, ordering objects centralize comparative judgments between systems, making it straightforward for architects to override expert-provided defaults. 
This separation reflects the reality that final deployment decisions rest with architects.

\begin{table*}[t]
\footnotesize
    \centering
    \begin{tabular}{r p{6in}}
    \toprule
    Type & Description  \\
    \midrule
    \texttt{Hardware} & A piece of network infrastructure (e.g. Cisco ASR9006 switch or xl170 HP server)\\
    \midrule
    \texttt{Objective} & A performance metric (e.g., latency or throughput) \\
    \texttt{System} & A deployable entity satisfying one or more {\tt Objectives} and imposing constraints on \texttt{Device}s and/or other {\tt Systems} (e.g., DCTCP~\cite{dctcp}) \\
    \texttt{Role} & Functionality that one or more \texttt{Systems} implement (e.g., congestion control algorithm) \\
    \texttt{Ordering} & A context-dependent ranking of {\tt System}s satisfying a given {\tt Role} for a given {\tt Objective} \\
    \midrule
    \texttt{Device} & A slot filled with a \texttt{Hardware} \\
    \texttt{Topology} & A flexible way to group devices with shared properties (e.g. rack, pod or datacenter) \\
    \texttt{Workload} & A description of an application deployed on a {\tt Topology}, including its properties, traffic matrix, and {\tt Objectives} (e.g., long scavenger flows in a datacenter that want high utilization) \\
    \texttt{Optimize} & A total ordering over \texttt{Workload}s' \texttt{Objectives} \\
    %\rahul{A total ordering of what the architect wants to optimize for} \\
    \bottomrule
    \end{tabular}
    \caption{\sysname types.}\label{t:types}
    \vspace{-10pt}
\end{table*}

\subsection{Interface}
\label{s:interface}
This section elaborates on the overview above to explain how \sysname can model a wide range of networking hardware, systems, objectives and workloads. We define a legible, extensible, and modular interface that supports encoding of a wide range of hardware, systems, objectives, and workloads as defined by both users and experts. Table~\ref{t:types} shows a summary of \sysname's types and their definitions.

\subsubsection{Describing Hardware}\hfill

\paragrapha{Hardware} 
A \texttt{Hardware}
\footnote{We implement all \sysname types as Python classes. For the rest of this section, we use monospaced font to refer to a Python type in the \sysname interface, e.g., \texttt{Hardware}, \texttt{Device}, etc.} 
instance is an individual piece of cloud infrastructure identified by a name and its specifications.
For example, the following describes a Cisco ASR 9006 switch:
\begin{lstlisting}
asr9006 = Hardware(
  id="CISCO_ASR9006", schema=switch_schema)
asr9006.set({
  "rack_units": 10, 
  "line_cards": 4,
  "ECN": True, 
  "INT": True,
  "port_bandwidth": Gbps(2048),
  "cross_section_bandwidth": Gbps(16384),
  ...})
\end{lstlisting}

\noindent
Other examples of {\tt Hardware} include an Nvidia V100 GPU or a Cisco 100 Gbit/s transceiver.

\noindent
\paragrapha{Device} A \texttt{Device} is a well-defined slot that a \texttt{Hardware} instance must fill. 
Each \texttt{Device} in a topology has a schema, which constrains what \texttt{Hardware} can occupy it.
The following shows the definition of the ``switch'' schema (the below listing elides some shared fields from the one above):
\begin{lstlisting}
switch_schema = HardwareSchema(
  cost = Entry(Real),
  cross_section_bandwidth = Entry(Real),
  ECN = Entry(Bool, default = False),
  QCN = Entry(Bool, default = False),
  QOS_Levels = Entry(Exhaustible, default = 1),
  ...)
\end{lstlisting}
\noindent
\texttt{Device} schemas can include both \emph{properties} and \emph{exhaustible resources}. Properties (either \texttt{Real} or \texttt{Bool}) expose attributes that \texttt{System}s can reference when expressing requirements or constraints. 
In contrast, exhaustible resources (\texttt{Exhaustible}) represent capacity-limited quantities that systems consume. \sysname tracks these resources' usage and ensures that output designs respect the available hardware limits.

\smallskip
\noindent\fbox{%
\centering
\parbox{0.95\columnwidth}{%
\textbf{Design Choice 1:} \texttt{Exhaustible} resources allow \sysname to perform limited quantitative reasoning when it makes sense to do, i.e., reasoning about numbers with resources but not attempting to derive performance metrics. 
}%
}
\smallskip

Schema fields can also optionally specify a default value; e.g., for optional hardware features.
For example, Timely~\cite{timely} uses one QoS level of a switch, whereas Homa~\cite{homa} can use as many as available. 
Thus, in a topology that uses the ASR 9006 switch above along with Timely and Homa for subsets of traffic, \sysname enforces that Timely occupies one dedicated QoS level and Homa occupies the remaining available levels.

\subsubsection{Describing Systems}\hfill 

\paragrapha{Objective} An \texttt{Objective} is a tag representing a metric.  Systems experts describe which objectives their \texttt{System} targets, and \texttt{Workload}s include which \texttt{Objective}s they benefit from.
The following shows the definition of two \texttt{Objective}s, ``fairness'' (lines 1--7) and ``app modification'' (line 8):
\begin{lstlisting}
fairness = Objective(
  id="fairness",
  granularities = [
    "per-flow"
    "per-application",
    "per-tenant",
  ])
app_modification = Objective(id="app_modification")
\end{lstlisting}
\texttt{Objective}s can define the \emph{granularity} at which they operate. For example, the fairness objective could refer to per-flow fairness (line 4), per-app fairness (line 5), or per-tenant fairness (line 6).
We note that \sysname does not reason about the meaning of each objective; rather, it simply matches objectives as tags to connect workload requirements to system benefits.

\vspace{3pt}
\noindent\fbox{%
\centering
\parbox{0.95\columnwidth}{%
\textbf{Design Choice 2:} Treating \texttt{Objectives} as tags keeps the design simple, but still enables complex reasoning. 
}%
}
\vspace{3pt}

\paragrapha{Role} \texttt{Systems} fulfill one or more \texttt{Role}s. 
\sysname specifies deploying one \texttt{System} per \texttt{Role}, but treats failing this constraint as a warning rather than a hard error.
The architect can also choose to disable a \texttt{Role} for a particular workload.

For example, we consider Annulus~\cite{annulus}, a system that identifies that competition between WAN and intra-datacenter traffic on the same queue causes high tail latencies. With \sysname, the Annulus experts define a \texttt{Role}, e.g., ``wan\_dc\_contention'' activated by this workload property.
The user can then choose to ignore the resulting warning or specify that \sysname should deploy a system filling this role.
Listing~\ref{l:annulus-role} shows how we encode the \texttt{Role} that the Annulus \texttt{System} fulfills. Lines 2--6 specify the condition required for this role to be active.%, i.e., for the issue to exist. 
\begin{listing}[t]
\begin{lstlisting}
def has_wan_dc_contention(workload):
  if wan_flows in workload.proprties:
    for other_workloads in workload.topology.workloads:
      if dc_flows in other_workloads.properties:
        return True
  return False
WAN_DC_Competition = Role(
  warning = "WAN-DC competition can cause high latency",
  activation_condition = has_wan_dc_contention)
\end{lstlisting}
\vspace{0.1in}
\caption{The definition of a \texttt{Role} activated when there is competition between WAN and DC flows. The need for that role is captured by the function \texttt{has\_wan\_dc\_competition}.}\label{l:annulus-role}
\vspace{-10pt}
\end{listing}
\paragrapha{System} A \texttt{System}\footnote{Sample \texttt{System} definitions are shown in the appendix.} is a unit of functionality which can either be deployed or not deployed, which satisfies one or more Objectives. 
Systems impose constraints on \texttt{Workload}s, other systems, and the workload's topology.

While \texttt{Role}s and thus system deployments are per-\texttt{Workload}, in many cases it is more efficient to use a single system deployment for multiple \texttt{Workload}s' \texttt{Roles}.
In some cases, reasoning about this resource sharing requires special encoding, since a system's resource consumption may scale differently with workloads.
For example, Timely~\cite{timely} requires 1 QoS level no matter how many hosts deploy it, but Andromeda~\cite{andromeda} requires more CPU cores with more workload.

\smallskip
\noindent\fbox{%
\centering
\parbox{0.95\columnwidth}{%
\textbf{Design Choice 3:} Decoupling system encodings from workloads and hardware allows the same system to satisfy multiple workloads and for multiple workloads to leverage the same hardware.
}%
}
\smallskip

\paragrapha{Ordering} \texttt{Ordering}s enable system experts and architects to rank systems in the same role with respect to a specific objective.
Of course, in general, choosing a system to deploy depends not only on objective metrics like performance but also subjective metrics like the architect's experience with each system.
\sysname's Orderings are partial by default, i.e., system experts can encode comparisons that they make against other systems of the same \texttt{Role}. 

Architects can add \texttt{Orderings} or override existing orderings based on their own benchmarks, preference, or experience (for objectives such as ease of deployment). 

\smallskip
\noindent\fbox{%
\centering
\parbox{0.95\columnwidth}{%
\textbf{Design Choice 4:} Using partial \texttt{Orderings} allows \sysname to capture how both architects and system experts reason about systems today.
}%
}
\smallskip

\begin{listing}[t]
\begin{lstlisting}
Workload(
  id = "ML_Training", 
  properties = [dc_flows, long_flows, incast]
  peak_cores_needed=3200, peak_bandwidth_needed=20,
  average_cores_needed=480, average_bandwidth_needed=10,
  ...)
\end{lstlisting}
\vspace{0.1in}
\caption{Description of an ML training \texttt{Workload}.}\label{l:workload_example}
\vspace{-15pt}
\end{listing}
\subsubsection{Architect Specifications}\hfill

\paragrapha{Topology} 
Architects express a network \texttt{Topology} using one or more \texttt{Device}s. Topologies are also composable. 

\paragrapha{Workload} A \texttt{Workload} describes functionality to run on the \texttt{Topology}.
As we described above, many modern systems improve performance in a specific context; \texttt{Workloads} thus encode constraints that \sysname can reason about in relation to systems' context, so that \sysname only uses relevant systems.
Listing~\ref{l:workload_example} shows an encoding of a machine learning training workload. The architect provides a name, ``ML\_Training,'' and 
describes the workload's properties.
Properties are tags (like objectives) that describe a workload. Experts can define \texttt{Role}s conditioned on these properties' presence. Finally, workloads use \texttt{Device}s' exhaustible resources similarly to \texttt{System}s. 

\paragrapha{Optimize} Architects can specify a total ordering over objectives, or 
optimize on numerical values such as the total hardware cost or total number of required cores.
While Optimize works as a strict priority, architects can also impose constraints for particular orderings (e.g., ``no worse than DCTCP'').

\paragrapha{Constraints}
Architects can also specify additional constraints to encode their institutional knowledge, e.g., organizations often wish to specify that redundant hardware devices should come from different vendors or that two specific hardware instances are incompatible, or logistical constraints, e.g., a cap on the number of RUs available on a WAN edge site or a service's maximum acceptable peak utilization.

\subsection{Constraint-Based Synthesis and Explainability}

\sysname formulates architecture design as a constraint-based optimization problem using Z3, enabling both feasibility checking and objective-driven synthesis. 
We encode each type described above in \S\ref{s:interface} in Z3 constraints (Appendix~\ref{s:encoding} describes this encoding).
\sysname configures Z3 to perform
lexicographic optimization according to the architect’s priorities. 
As a result, \sysname produces a deployment that satisfies all constraints while optimizing for the specified objectives, or report infeasibility when no such configuration exists. 

Beyond synthesis, we design \sysname to support explainability, which we define as the question ``\emph{what prevents an alternative, user-preferred choice from appearing in an optimal design?}''
When a user identifies a system $b$ that they expected to be chosen instead of system $a$, an LLM frontend to \sysname attempts to redesign a configuration in which $b$ replaces $a$. 
To allow users to have maximum control over the flexibility of the new configuration compared to the older one, it permits users to explicitly specify any subset of choices, and checks whether a feasible configuration exists under these modified constraints.
If the solver then returns a feasible configuration, the system presents it alongside the original, making explicit the trade-offs introduced, since by design improving on one objective must degrade another.
If no feasible configuration exists, \sysname extracts the UNSAT core and its LLM frontend translates it into a human-readable explanation identifying the conflicting constraints.
To improve the explanation's clarity, \sysname recursively decomposes conjunctions and implications from an UNSAT core into smaller atomic clauses, each corresponding to a specific dependency or requirement. 
It then re-solves the refined constraint set, restricted to the original UNSAT core, to isolate the minimal subset responsible for infeasibility. 
Although this requires an additional solver step, it operates on a much smaller input and substantially improves the clarity and usefulness of the resulting explanations.

By definition, \sysname excludes a system from a configuration for one of three underlying reasons: \emph{workload mismatch}, \emph{insufficient inventory}, or \emph{system incompatibility}. A workload mismatch occurs when the system is incompatible with the workload’s properties, orderings, or explicitly specified objectives. Insufficient inventory arises when no available hardware configuration can satisfy the system's resource requirements, either in isolation or in combination with other components. System incompatibility captures conflicts between the candidate system and other systems in the design, including those fixed by the architect or implied by feasible alternatives. These categories provide actionable guidance, helping users refine objectives, revise workload specifications, or expand available resources. A formal characterization that restricts possible explanations is given in Appendix~\ref{sec:explain_formal}.

\sysname allows experts to attach natural language explanations to each constraint specified for a system or hardware component. These annotations are used to improve the clarity and interpretability of the generated explanations. Additionally, the structured output is passed to an LLM to produce a concise, user-friendly summary.

\subsection{Limitations}
\sysname's embedding cannot capture some potentially useful points in the design space, and we discuss three such points next. We leave the design of an embedding that captures these aspects to future work.

First, the topology is fixed and \sysname does not consider changing it.
\sysname thus can't analyze changes to the topology that can improve application performance or other objectives, such as those described in past work such as Jellyfish~\cite{jellyfish}, Opera~\cite{expander-graphs}, and SiP-ML~\cite{sip-ml}. 
Second, workloads currently specify an exact number of devices they run on. \sysname does not attempt to analyze elastic scaling behavior, i.e., whether the workload would be better off running on a larger or smaller set of devices.
Third, \sysname encodes a workload's properties and objectives statically and does not represent them in Z3. This means that a system cannot alter them. This is relevant in cases where deploying a system changes the nature of an application's impact on the network. 
For example, consider an ML training application: using a parameter server will have a different impact on the network than using all-reduce, but \sysname cannot model such derivative changes to workloads.
Instead, currently the architect must separately consider separate workload instances for each of these options, and \sysname can then reason about these options separately.

\section{Evaluation}\label{s:casestudies}

We evaluate \sysname through two case studies that reflect common architectural decision-making scenarios. 
The first adopts the perspective of an application architect selecting a deployment stack for a microservices-based application, including components such as the orchestrator, autoscaler, service mesh, container runtime, RPC library, and a storage backend for monitoring data. 
The second considers a datacenter network architect tasked with provisioning infrastructure for a set of applications, beginning with an AI workload and progressively incorporating additional requirements.

Each case study highlights a different aspect of \sysname, and the significant difference between the case studies shows \sysname's ability to adapt to different design tasks. 
In the first, we compare the synthesized design of \sysname with a one-shot output of an LLM. The structure of this problem allows for both qualitative and quantitative comparisons, illustrating differences in how each approach navigates the design space. In the second case study, we focus on explainability, examining how \sysname surfaces the constraints and trade-offs underlying its decisions in the presence of complex interdependent design choices.

To support these evaluations, we encoded over fifty systems in \sysname across a diverse set of roles, including network stack, congestion control, network monitoring, firewalls, virtual switches, load balancers, transport protocols, service mesh, and orchestrators. Each system is specified by domain experts, ensuring the correctness of its constraints. We also model hardware across multiple device types (e.g., compute nodes, switches, routers) using publicly available specifications, including configurations derived from CloudLab~\cite{duplyakin2019design}, where parts of our evaluation are validated. In total, we implemented \sysname and the case studies in \textasciitilde 3,000 lines of Python.

Through this evaluation, our aim is to answer the following concrete questions: 
\begin{itemize}[noitemsep,topsep=0pt,leftmargin=*]
\item Can \sysname identify configurations that better satisfy stated objectives than those produced by LLMs alone?
\item How much performance improvements can be expected when \sysname is used compared to an LLM alone?
\item How well does \sysname capture non-obvious cross-component interactions that impact performance and cost? 
\item Can its explanations help users understand and refine their design decisions when expectations and outcomes diverge?
\end{itemize}

\begin{figure}[!t]
    \centering
    \includegraphics[width=0.6\linewidth]{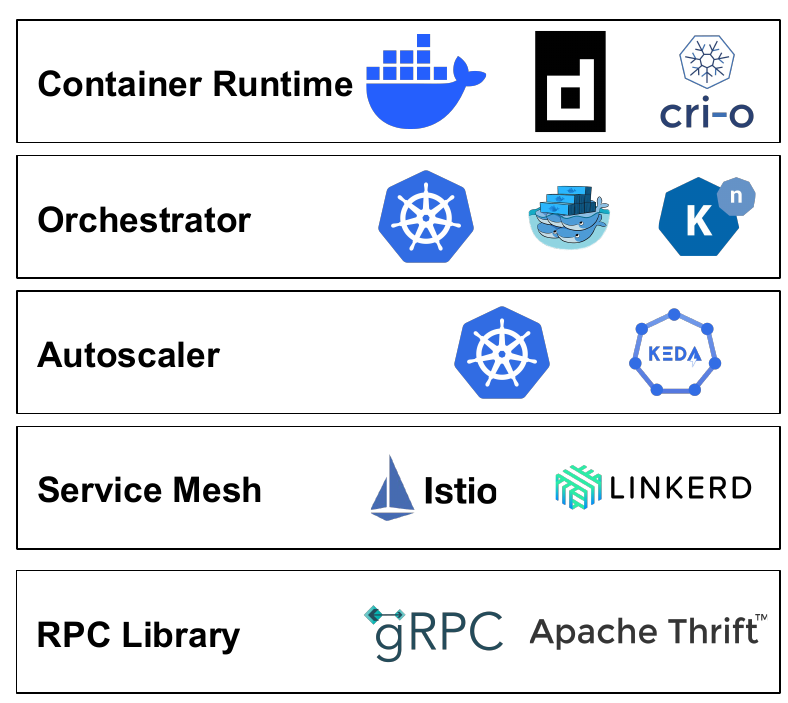}
    \caption{An illustration of design options in the cloud-native platforms stack.}
    \label{f:cs1_design_choices}
    \vspace{-0.1in}
\end{figure}

\subsection{\sysname for Application Architecture Design}
\label{sec:case_study_1}

\paragrapha{Problem context}  We consider an architect designing the deployment configuration for the Hotel Reservation application from DeathStarBench~\cite{deathstarbench}. The goal is to leverage modern cloud-native frameworks while balancing two objectives: ease of deployment and low end-to-end latency. We define ease of deployment as requiring minimal effort beyond standard documentation for setup, operation, and debugging. Low end-to-end latency captures the additional overhead introduced by the deployment stack along the request path.

The application consists of 18 microservices: a frontend service, a Jaeger service for tracing, nine backend storage services, and seven intermediate services that communicate via gRPC. Figure~\ref{f:cs1_design_choices} summarizes a subset of the design choices available to the architect, all of which are encoded in \sysname. 
For comparison, we also generate deployment configurations using two LLM services, ChatGPT and Gemini. 
We observed no significant differences between their outputs and therefore focus on ChatGPT. 
Notably, while \sysname selects from a fixed library of encoded systems, the LLM is not constrained in this way and can propose any arbitrary configuration.

\begin{table}[t]
\footnotesize
    \centering
    \begin{tabular}{c c c}
        \toprule
        \textbf{Role} & \textbf{\sysname} & \textbf{LLM} \\
        \midrule
        \textbf{Container Runtime} & Containerd & Docker \\
        \textbf{Orchestrator} & Kubernetes & Kubernetes / Knative \\
        \textbf{Autoscaler} & Kubernetes HPA & Kubernetes / Knative \\
        \textbf{Service Mesh} & Linkerd & Istio (sidecar mode) \\
        %\textbf{Storage Backend} & Elasticsearch & Elasticsearch \\
        %\hline
        \textbf{RPC} & gRPC & gRPC \\
        \bottomrule
    \end{tabular}
    \caption{Configuration with Ease of Deployment as top priority}
    \label{t:eod-config}
    \vspace{-0.2in}
\end{table}
\begin{table}[t]
\footnotesize
    \centering
    \begin{tabular}{c c c}
    \toprule
    \textbf{Role} & \textbf{\sysname} & \textbf{LLM} \\
    \midrule
    \textbf{Container Runtime} & Containerd & Docker \\
    \textbf{Orchestrator} & Kubernetes & Kubernetes \\
    \textbf{Autoscaler} & Kubernetes HPA & Kubernetes HPA \\
    \textbf{Service Mesh} & Istio (ambient mode) & Istio (ambient mode) \\
    %\textbf{Storage Backend} & Cassandra & Elasticsearch \\
    %\hline
    \textbf{RPC} & gRPC & gRPC \\
    \bottomrule
    \end{tabular}
    \caption{Configuration with Latency as top priority}
    \label{t:latency-config}
    \vspace{-0.2in}
\end{table}

\paragrapha{Designing for ease of deployment} We first ask both \sysname and the LLM to synthesize a deployment that prioritizes ease of deployment, with latency as a secondary objective. Table~\ref{t:eod-config} summarizes the resulting configurations. The two approaches agree on some roles (e.g., orchestrator and RPC library) but diverge on others, including the container runtime, autoscaler, and service mesh. We examine these differences and their implications.

\smallskip
\noindent\fbox{%
\centering
\parbox{0.95\columnwidth}{%
\textbf{Takeaway:} LLM services we evaluated exhibited biases toward existing documentation and common patterns.
}%
}
\smallskip

\noindent
To understand the LLM’s choices, we note that the DeathStarBench documentation provides recommended configurations for several components (all except the autoscaler), and the LLM largely reproduces these recommendations. However, these defaults are not aligned with the stated objective of ease of deployment.
We also observe ``stickiness'' to these documented configurations, even when they are outdated or suboptimal for the objective. For example, the LLM selects Docker as the container runtime as the documentation recommends. 
In practice, Kubernetes has deprecated direct Docker integration in favor of Container Runtime Interface (CRI)-compliant runtimes such as containerd and CRI-O. 
While Docker-based setups remain functional, they introduce additional configuration complexity. In contrast, \sysname selects containerd, which better aligns with current Kubernetes practices and simplifies deployment.

Second, the LLM's output fixated on detailed examples in documentation. 
In this case, it proposes a hybrid deployment where a subset of services is managed using Knative and another subset using Kubernetes, motivated by references to serverless deployment in the benchmark. 
However, serverless operation is not an explicit objective here. 
A simpler and more consistent design is to use Kubernetes with its built-in Horizontal Pod Autoscaler (HPA) across all services. Similarly, the LLM selects Istio, the service mesh referenced in the documentation, despite its higher latency overhead in default configurations. \sysname instead selects Kubernetes with HPA for orchestration and autoscaling, and Linkerd as a service mesh, resulting in a simpler and lower-latency deployment.

We note that iterative prompting can improve LLM-generated designs. However, doing so effectively requires substantial expertise: the architect must recognize and correct the LLM's suboptimal suggestions or validate them through deployment and debugging, both of which are time-consuming. 
Moreover, LLM outputs are inherently nondeterministic and sensitive to prompt context, making them less reliable for systematic design exploration.

\paragrapha{Designing for low latency} We next change the objective to prioritize low latency over ease of deployment. Table~\ref{t:latency-config} compares the resulting configurations. The two approaches now agree on all roles except the container runtime. Importantly, both select Istio configured in ambient mode, which offers a favorable latency profile compared to its default deployment. 

This shift highlights \sysname's ability to adapt its choices based on changing priorities: once ease of deployment is deprioritized, the additional complexity of Istio in ambient mode becomes acceptable in exchange for improved latency. In contrast, while the LLM partially adjusts its design (e.g., selecting Istio in ambient mode), some of its earlier choices persist. In particular, it continues to select Docker as a container runtime, reflecting ``stickiness'' toward default or documented configurations, even when they are no longer aligned with the primary objective.

\begin{figure}[!t]
    \centering
    \includegraphics[width=0.9\linewidth]{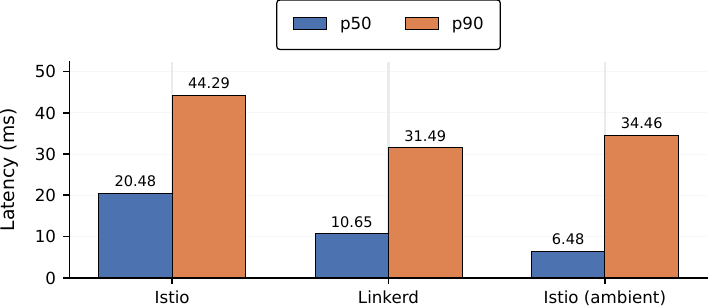}
    \caption{Median and p90 end-to-end latency of requests made the Hotel Reservation application under different service mesh configurations.}
    \label{f:cs1_latency}
    \vspace{-0.3in}
\end{figure}

\paragrapha{A quantitative validation} We evaluate three configurations derived from \sysname’s output, varying only the choice of service mesh to isolate its impact on latency. This experiment focuses on the primary quantitative metric of interest: end-to-end latency. We deploy the Hotel Reservation application on ten c220g1 nodes on CloudLab~\cite{duplyakin2019design}, each equipped with 16-core Intel E5-2630 CPUs running Ubuntu 22.04 LTS. One node serves as both the controller and client, while the remaining nodes host the application. The client generates requests at a rate of 90 RPS.

Figure~\ref{f:cs1_latency} shows that Istio in ambient mode and Linkerd achieve comparable tail latency, both outperforming Istio’s default configuration. However, Istio in ambient mode provides significantly better median latency than Linkerd. Overall, these results validate \sysname’s ability to select configurations that align with the specified latency objective and help highlight the LLM's problematic decision when it picked Istio in sidecar mode.

\begin{figure}[t]
\centering
\vspace{-0.1in}
\includegraphics[width=0.9\linewidth]{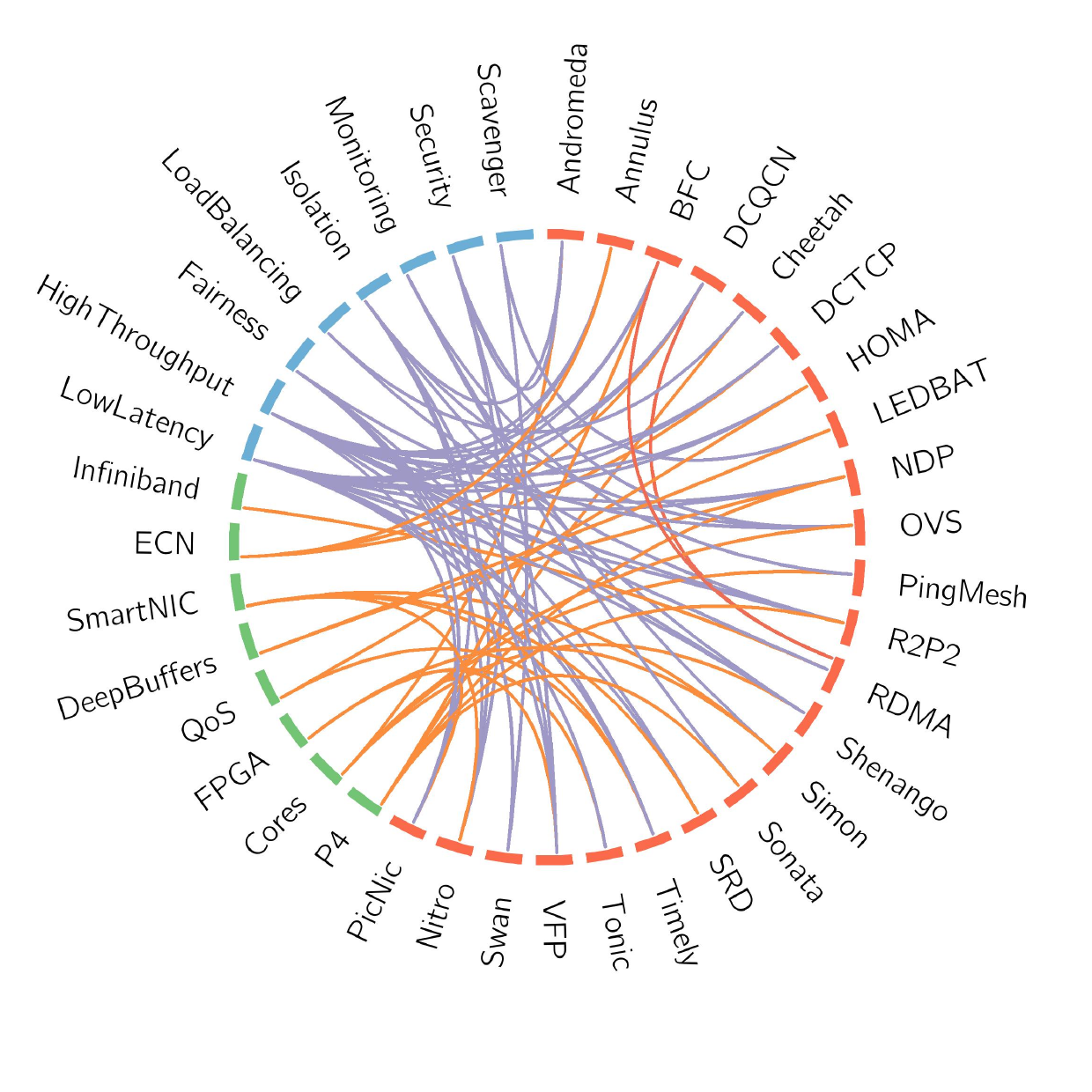}
\vspace{-0.4in}
\caption{Visual representation of the hardware requirements (green) of some of the Systems (red) we encoded and the objectives (blue) they fulfill.}
\label{fig:rel}
\vspace{-0.2in}
\end{figure}

\begin{listing}[t]
\begin{lstlisting}
pod = create_pod("pod1", 
  num_racks = 2, num_routers = 1, num_cpus = 7)
racks = pod.get_children(
  devicegroup_type = DEVICEGROUP_TYPE.RACK)
ML_Training_Workload = Workload(
  "ML_Training", 
  racks[0], 
  [dc_flows, long_flows, ml_training], 
  [latency, throughput, ease_of_deployment, monitoring], 
  network_load=9)
  
ML_Training_Workload.set_performance_bound(
  objective = load_balancing, atleast = PLB)

Optimize(ML_Training_Workload, latency, 1)
Optimize(ML_Training_Workload, throughput, 2)
Optimize(ML_Training_Workload, ease_of_deployment, 3)
---
Time taken by z3 to solve: 0.9041211605072021
** Systems deployed **
ML_Training -
1. cpu_sched role: ZygOS
2. cca role: DCQCN
3. virtual_switch role: ANDROMEDA
4. transport role: RDMA
5. load_balancer role: PLB
6. Monitor role: Sonata

** Other values **
Total cost = 10450

** Hardware assignments **
pod1IN-POD_1IN-RACK_0_DEVICE_TYPE.COMPUTE = 4Cores_highFPGA
pod1IN-POD_1IN-RACK_3_DEVICE_TYPE.COMPUTE = 4Cores_highFPGA
pod1IN-POD_1IN-RACK_5_DEVICE_TYPE.COMPUTE = 8Cores_highFPGA
pod1IN-POD_0_0_DEVICE_TYPE.LINK = 80GLink
pod1IN-POD_0IN-RACK_DEVICE_TYPE.ROUTER = TofinoV1Router
pod1IN-POD_1IN-RACK_2_DEVICE_TYPE.COMPUTE = 4Cores_rdma
\end{lstlisting}
\caption{Description of the ML training workload with \sysname, and \sysname's output.}\label{l:ml-training-workload}
\vspace{-10pt}
\end{listing}

\subsection{\sysname for Datacenter Network Design}
\label{sec:case_study_2}

\paragraph{Problem context} Datacenter network design is a large and rapidly evolving search space, driven by continuous innovation across the networking stack. Design decisions are highly inter-dependent: the choice of one system constrains the set of viable options for other roles and may impose specific hardware requirements. Figure~\ref{fig:rel} illustrates a subset of these interactions as encoded in \sysname.

In this case study, we consider the problem of designing a network to support a set of applications. We begin with an initial scenario in which the architect optimizes for a single application, and then progressively introduce additional requirements. This setting allows us to demonstrate how \sysname adapts the design in response to evolving constraints while maintaining a consistent and explainable reasoning process.

\noindent\textbf{Single workload.} We begin by using \sysname to design a cluster for a single low-latency ML training workload (Listing~\ref{l:ml-training-workload}). The topology (lines 1--4) consists of a single pod with two racks, each containing seven compute nodes connected via ToR switches and routers. Links are defined between each compute node and its rack. The workload specification (lines 5--10) follows the standard format, with an additional requirement that packet load balancing be at least as good as PLB (lines 12--13). The architect prioritizes latency, throughput, cost, and monitoring, in that order (lines 15--17).

\smallskip
\noindent\fbox{%
\centering
\parbox{0.95\columnwidth}{%
\textbf{Takeaway:} \sysname can capture and \emph{explain} non-trivial interactions between components.
}%
}
\smallskip

\sysname's output (starting at line 19) reflects non-trivial interactions between systems and hardware constraints. 
For example, it selects Andromeda for the virtual switch role, citing comparable latency and throughput to alternatives while requiring fewer resources and being easier to deploy. 
For load balancing, PLB is selected since sufficient CPU resources are available. 
ZygOS (CPU scheduler) and RDMA (transport) are chosen to optimize latency. 
However, BFC congestion control is not selected despite its performance benefits, because the available hardware does not support links beyond 100Gbps; instead, \sysname selects DCQCN, which imposes only QoS-related constraints.

\begin{listing}[t]
\begin{lstlisting}
from systems.conga import CONGA
from systems.packetspray import PACKETSPRAY
from systems.plb import PLB
from systems.ecmp import ECMP

Ordering(load_balancing, PACKETSPRAY, better_than = CONGA)
Ordering(load_balancing, CONGA, same_as = PLB)
Ordering(load_balancing, PLB, better_than = ECMP)

Ordering(latency, PACKETSPRAY, better_than = CONGA)
Ordering(latency, CONGA, better_than = ECMP)
Ordering(latency, CONGA, same_as = PLB)

Ordering(ease_of_deployment, PLB, better_than = CONGA)
Ordering(ease_of_deployment, ECMP, better_than = CONGA)
Ordering(ease_of_deployment, 
  PACKETSPRAY, better_than = CONGA)
Ordering(ease_of_deployment, 
  ECMP, better_than = PACKETSPRAY)
\end{lstlisting}
\caption{Ordering for Load balancers}\label{l:load-balance-ordering}
\vspace{-10pt}
\end{listing}

\begin{table}[t]
\centering
\footnotesize
\begin{tabular}{p{1.55cm} p{6.0cm}}
\toprule
\textbf{Item} & \textbf{Constraint / implication} \\
\midrule
\textsc{RDMA} & Requires every compute device to satisfy \texttt{RDMA = True}. \\
\textsc{PacketSpray} & Requires every compute device to satisfy \texttt{NIC\_Reorder\_Buffer > 20}. \\
Hardware & No hardware option satisfies both \texttt{RDMA = True} and \texttt{NIC\_Reorder\_Buffer > 20}. Available devices either have \texttt{RDMA = False}, or reorder buffer values of only 10 or 20. \\
Result & \textsc{RDMA} $\wedge$ \textsc{PacketSpray} is unsatisfiable; hence \textsc{PacketSpray} is not selected. \\
\bottomrule
\end{tabular}
\caption{Why \textsc{PacketSpray} is not selected for \textsc{ML\_Training}.}
\label{tab:packetspray-rdma}
\vspace{-0.2in}
\end{table}

\sysname selects the top-ranked system for all roles other than load balancing, for which  
PacketSpray ranks highest (Listing~\ref{l:load-balance-ordering}).
We use 
\sysname's explainability (Table~\ref{tab:packetspray-rdma}) to understand this decision.
The solver reports the configuration as UNSAT, identifying a conflict between PacketSpray and RDMA. Specifically, RDMA requires all compute devices to support RDMA-capable NICs, while PacketSpray imposes constraints on NIC reorder buffer sizes. 
The available hardware does not satisfy both requirements simultaneously. As a result, no feasible configuration exists that includes both PacketSpray and RDMA, and \sysname selects PLB as the next-best option. This example highlights how \sysname captures subtle cross-layer dependencies and uses them to rule out otherwise higher-ranked choices. Note that Table~\ref{tab:packetspray-rdma} is the final output of \sysname the LLM frontend summarizes the filtered and labeled UNSAT core. 
We show this intermediate output in the appendix.

\noindent\textbf{Adding Frontend workload.}
The architect next adds a frontend workload to the deployment and uses \sysname to optimize performance under the same topology (i.e., within the existing rack space). Listing~\ref{l:frontend-workload-def} shows the new workload, deployed on \texttt{racks[0]}. For this workload, \sysname selects Conga as the load balancer, since ease of deployment is not an objective. To support the combined requirements of Andromeda, Simon, and RDMA, the hardware is upgraded to 16 cores per compute node, increasing the total cost.

The architect then decides to prioritize ease of deployment for the virtual switch, based on team expertise. Such scenarios highlight a key strength of \sysname: even a single change in the preferences of the architect can trigger cascading adjustments across the design. In this case, \sysname replaces the monitoring system with Pingmesh to free NIC resources, since Simon consumes NIC capacity. It further determines that the only hardware configuration in its inventory with sufficient FPGA capacity and CPU cores to support both workloads does not support RDMA; as a result, it replaces RDMA with Pony in both workloads. 

We present more case studies 
in Appendix~\ref{app:case_studies}.

\begin{listing}[t]
\begin{lstlisting}
FrontEnd_Workload = Workload("FrontEnd", racks[0], [internet_flows, high_priority, long_flows], [latency, throughput, fairness, security, fault_tolerance, application_modification, ease_of_deployment], compute_load=10 ,network_load=20)
Optimize(FrontEnd_Workload, latency, 1)
Optimize(FrontEnd_Workload, application_modification, 2)
Optimize(FrontEnd_Workload, monitoring, 3)
\end{lstlisting}
\vspace{0.1in}
\caption{Frontend Workload by user}\label{l:frontend-workload-def}
\vspace{-0.2in}
\end{listing}

\section{Related Work}\label{s:related}

Prior approaches to modelling network architectures have generally focused on heavyweight verification. 
For example, work in dataplane verification~\cite{Kazemian2012HSA,Khurshid2013VeriFlow,Kazemian2013HSA2,Fogel2015Batfish,Beckett2017minesweeper, anteater,panda-reachability-nsdi17,Prabhu2020Plankton,Yuan2020NetSMC,yousefi2020liveness,Jiang2020PCF,Dobrescu2014DPDKVR,Zaostrovnykh2019nfv,kim2015kinetic} has used this technique to detect misconfigurations or determine whether some network invariant holds. 
While, like \sysname, some of these techniques use SMT solvers, their goals are lower-level than \sysname's: to verify that the implementations of two network systems are inter-operable by inspecting their source code.
Rather, \sysname does not attempt to understand systems' implementations, or even their abstract behavior. Instead, it provides an interface for system experts and network architects to communicate their domain-specific knowledge to each other using machine-readable rules.

An alternative approach to building network designs is intent-based specification~\cite{Soul2014Merlin, Prakash2015PGA, Heorhiadi2018Intent}. With this approach, network traffic engineers write specifications for how the network should react to dynamic events such as load changes or failures. Special compilers then produce verified code that runs in network elements to enforce the appropriate policy.
Rather than replacing existing systems with ones produced by these compilers, \sysname instead reasons about them.

Robotron~\cite{robotron} and MALT~\cite{malt} manage hardware inventory and complex infrastructure by encoding hardware instances and their configurations (e.g., IP addresses, wiring, etc.), in a machine-readable way. 
\sysname does not reason about the specifics of a network topology; rather, it focuses on network systems.
This enables higher-level reasoning (e.g., enforcing constraints on congestion control algorithms), but is not as well suited for network lifecycle management.

Meanwhile, QARC~\cite{Subramanian2020QARC} and Chang et al.~\cite{Chang2017demand} specifically seek to determine whether a given topology can accommodate an input traffic workload without becoming overloaded. These approach are more accurate than the simple one \sysname takes, and could be integrated with \sysname in future versions.

Finally, early position papers from Liu et al.~\cite{lie2020selfdriving}, Gong et al.~\cite{maria-hotnets-fm-llm}, and Sharma et al.~\cite{prosper} propose various combinations of LLM and formal models to solve a variety of networking challenges from discovering bugs in controllers to more accurately modeling individual components.
Insights from these could be incorporated in \sysname.

\section{Conclusion}

Designing networked system architectures requires reasoning over a large space of interacting components, constraints, and objectives—a task that goes beyond generating plausible configurations. Our evaluation shows that while LLMs are useful for information retrieval and high-level suggestions, they are not reliable for end-to-end architectural reasoning. While iterative validation could improve their outputs, it is frequently impractical due to the cost and scale of required simulations or deployments, and infeasible when reasoning about hardware-dependent trade-offs.

\sysname demonstrates an alternative approach based on structured representations and lightweight formal reasoning. By encoding architecturally significant properties as constraints and optimizing over qualitative objectives, it systematically synthesizes feasible designs while capturing cross-layer interactions that are easy to overlook. Its explainability mechanism further enables users to understand trade-offs and refine designs in a principled manner.

More broadly, our findings suggest that the community should move toward building shared, machine-readable representations of system knowledge—capturing constraints, capabilities, and trade-offs—and combining them with formal reasoning tools. LLMs can play a valuable role in extracting and interfacing with this knowledge, but should be grounded in such structured representations rather than used as standalone design agents.

This work does not raise any ethical issues.
%\pagebreak

\bibliographystyle{plain}

\begin{small}
\bibliography{refs}
\end{small}
\appendix

\section{Constraint Encoding and Architecture Design Synthesis}\label{s:encoding}

\begin{listing}[!t]
\begin{lstlisting}
def pingmesh_constraints(workload):
  devices = workload.get_deployed_devices()
  computes = devices.get(DEVICE_TYPE.COMPUTES)
  constraints = []
  for compute in computes:
    cores_per_server = compute.allocate("cores", PINGMESH_CPU_FACTOR * len(computes))
    constraints.append(cores_per_server)
  return constraints
PingMesh = System(
  id = "PingMesh"
  role = MONITOR
  deployment_constraints = pingmesh_constraints,
  solves = monitoring.capture_delays
)
\end{lstlisting}
\vspace{0.1in}
\caption{Description of PingMesh.}\label{l:pingmesh}
\vspace{-10pt}
\end{listing}

We use PingMesh, a network monitoring system,  as a running example in explaining our encoding.
Listing~\ref{l:pingmesh} shows an example description of PingMesh~\cite{pingmesh}, . 
Lines 1-10 describe the constraints that \sysname should enforce if PingMesh is deployed. Lines 5-7 state that PingMesh uses CPU resources based on the total number of servers it is deployed on; we computed the value of \texttt{PINGMESH\_CPU\_FACTOR} to be close to 8e-5 based on the paper's description. Finally, line 11 instantiates the system with the \texttt{Role} ``Monitor''.

\sysname uses Z3 as its reasoning engine to perform two functions: (i) model checking, ensuring that all performance objectives and system requirements are met, and (ii) optimization, e.g., picking the deployment that best meets the architect's goals. This requires encoding the input described in \S\ref{s:design} into SMT formulas that Z3 can solve. We summarize the encoding as:
\[(Constraints_{Devices} \wedge Constraints_{Roles} \wedge Constraints_{Orderings})\]
The solver then finds a viable deployment of systems and hardware that satisfy these constraints, or returns an error if this is not possible.

\paragrapha{Device Constraints} Every device in the topology should be filled with a hardware, and that hardware's schema must match the device's:
\begin{align*}
\forall \textnormal{ devices }& d: 
\exists \textnormal{ hardware }  h \textnormal{ s.t.:} &\\
&\hspace{-0.2in}(\texttt{d.hardware\_id} == \texttt{h.id } \wedge  
\texttt{ d.type} == \texttt{h.type} )\\
& \hspace{-0.1in} \implies       (\texttt{d.configuration} == \texttt{h.configuration}) )
\end{align*}

\noindent
%\changebars{
Further, each device's constrained slots should have enough capacity to accommodate the aggregate demand of the systems and workloads that utilize it.
%}{Further, the selected hardware should have enough capacity in its constrained slot to fit all the systems that should be deployed on it. }
\begin{align*}
&\forall \textnormal{ devices } d:  \forall \textnormal{ slots } s \in d: \\
&\texttt{d.configuration[s]} \geq \\
&\hspace{10pt}\sum_{\forall sys \in \texttt{deployed on } d} \texttt{d.constrained\_slot[s][sys]}
\end{align*}
\sysname applies similar requirements for workloads' demands.
Recall that systems' and workloads' definitions specify their demands (e.g., Line 7 in Listing~\ref{l:pingmesh} and Line 2 in Listing~\ref{l:workload_example}).
\paragrapha{Role Constraints} Roles are activated based on workload properties (e.g., the role defined in Listing~\ref{l:annulus-role}). Further, architects can explicitly state whether a role should be considered or not. We capture that dependency through the following constraint: 
\begin{align*}
\forall& \textnormal{ workloads } w: \forall \textnormal { roles } r: \\
&\hspace{0.1in}\texttt{r.considered } \wedge \texttt{ r.apply(w) } \implies \texttt{ r.enabled }
\end{align*}
Roles determine which systems should be considered for deployment. 
A role can be exclusively held by a single system or shared by multiple systems ($\veebar$ represents exclusive or). 
% \forall \textnormal{ systems } s:
\begin{align*}
&\forall \textnormal{ workloads } w: \forall  \textnormal{ roles } r: \\
&(\texttt{r.enabled} \wedge \neg\texttt{r.is\_exclusive}) \implies \\
&\hspace{10pt}\bigvee_{s \in \textnormal{ systems}}(\texttt{w.deployed(s) } \wedge \texttt{ s.role == r})\\
&(\texttt{r.enabled} \wedge \texttt{r.is\_exclusive}) \implies \\
&\hspace{10pt}\ubigvee_{s \in \textnormal{ systems}}(\texttt{w.deployed(s) } \wedge \texttt{ s.role == r})
\end{align*}
A deployed system applies its constraints to the workload (e.g.,  
Line 1 in Listing~\ref{l:pingmesh}).

\begin{align*}
&\forall \textnormal{ workloads } w: \forall \textnormal{ systems } s:  \\
&\hspace{10pt}\texttt{w.deployed(s)} \implies \texttt{s.constraints(w)}
\end{align*}
Recall that the definition of a workload includes a reference to the devices on which it will be deployed. Thus, the system can apply constraints that consider both workloads as well as devices and systems on that workload.

\paragrapha{Orderings and Optimizations} Orderings are already in constraint form, and we encode them as-is. We encode Optimizations by passing them to \texttt{z3.Solver.Maximize()} in the given priority order.

\paragraph{Modularity of constraints}
The architect can run into situations where they constrain their design to the point that there is no feasible solution. For example, expressing that no hardware or system should change but adding a new workload might not be feasible. In these cases, the architect can iteratively relax the constraints to find a feasible solution. 
An example relaxed query might express willingness to change the hardware or systems, but a preference to use already-deployed solutions. To support this, \sysname supports marking constraints as optional.

\section{Explainability}\label{s:explainability}
The case-studies in Section \ref{s:casestudies} present scenarios where the user benefits from iterative improvement of their workload definition and understanding of the systems encoded. We recognize the importance of this feedback loop, with the tool able to provide reasons, through constraints, over selection of particular systems when it does not match the user's expectations. Not only does this avoid the tool becoming a black-box, but also, due to its inherent formal structure, allows the user to pinpoint the precise constraints and dependencies that influence the optimal result presented by \sysname. 
Given that the user is prompting for an explanation for an already existing optimal configuration (output of the main query), our explainability flow attempts to fix as many choices from it, only relaxing the choice that the user specifies explicitly. In other words, the framework answers the following question: \textbf{Given an optimal configuration C, if instead of system \textit{a}, system \textit{b} was picked, and there were only a subset of C, C$'$ that was allowed to change, is an optimal configuration possible, and if not, why?}

\begin{algorithm}[t]
\small
\caption{Explainability extended procedure}
\label{alg:explainability}
\begin{algorithmic}[1]
     \Type \\ Explained Output = Possible Output Class | [Constraints]
     \Input \\
     Hardware Choices: $H$ \\
     System Choices: $S$ \\
     Ordering: $O$ \\
     Objective: $obj$ \\
     Flexible Roles: $R_f$
     \Output \\
     Explained Output: $E$
     \Procedure{Explain}{$H$, $S$, $O$, $obj$, $R_f$}
     \State $T$ $\gets$ \Call{GetPrioritySystem}{$S$, $O$, $obj$}
     \If{$T = \emptyset$} 
        \State \Return $\emptyset$
     \Else 
        \Comment{$T$ is a higher priority system not picked}
        \State \Call{FixChoice}{$T$}
        \State \Call{FixChoices}{$(H \cup S) - R_f$} 
        \State $Output_{new}$ $\gets$ \Call{Solver}{}
        \If{$Output_{new}$ == UNSAT}
            \State $Core$ $\gets$ \Call{GetCore}{}
            \State $Constraints_{UNSAT}$ $\gets$ \Call{BreakDownCore}{$Core$}
            \State \Call{Solver}{$Constraints_{UNSAT}$}
            \State $NewCore$ $\gets$ \Call{GetCore}{}
            \State \Return $NewCore$
        \Else
            \State \Return $Output_{new}$
        \EndIf
     \EndIf
     \EndProcedure
\end{algorithmic}
\end{algorithm}

\subsection{Classification of explainability results}

Explainability provides an interface for users to understand why certain systems are not selected by exposing the constraints that prevent their inclusion. We observe that explanations fall into three primary categories: \textbf{workload mismatch}, \textbf{objective misalignment}, \textbf{insufficient inventory}, and \textbf{system incompatibility}, with \textbf{non-optimal cost} arising when feasible alternatives are strictly worse under the objective ordering. 

A \textbf{workload mismatch} occurs when a system is incompatible with the workload’s properties or topology. \textbf{Objective misalignment} is when there is a mismatch between a system's intended objectives and the workload's specified objectives. \textbf{Insufficient inventory} arises when no available hardware configuration can satisfy the system's resource requirements, either independently or in combination with other choices. \textbf{System incompatibility} captures conflicts between the candidate system and other systems in the design, including those fixed by the architect or implied by feasible alternatives. 
These categories provide actionable guidance, enabling users to refine workload specifications, adjust system choices, or expand available resources in subsequent iterations.

\subsection{Completeness of Explainability Classification} 
\label{sec:explain_formal}
This section formalizes the space of explanations produced by \sysname and shows that the reasons for excluding a system from an optimal configuration fall into a small, well-defined set of categories.

\paragraph{Setup.}
Let $C$ denote an optimal configuration consisting of assignments $r_1!:!c_1, r_2!:!c_2, \dots, r_n!:!c_n$ for $n$ roles, where $r_x$ denotes role $x$ and $c_x$ is the system selected to fulfill it. By definition, all assignments in $C$ satisfy the workload specification $W$, which includes architect-defined constraints, orderings, and optimization objectives.

Under the explainability framework, we consider an alternative system $S \notin C$ for some role, while allowing a subset of choices $C' \subseteq C$ to vary. Formally, for each $r_y \in C'$, the assignment $c_y$ may be replaced, whereas for each $r_z \in C \setminus C'$, the assignment $c_z$ is fixed.

\paragraph{Overview.}
A system $S$ is not selected in \sysname's chosen configuration for workload $W$ if and only if at least one of the following conditions holds:
\begin{enumerate}
\item \textbf{Workload mismatch:} $S$ is incompatible with $W$ (e.g., due to workload orderings or explicit constraints);
\item \textbf{Objective misalignment:} the objectives satisfied by $S$ are not aligned with those specified in $W$;
\item \textbf{Insufficient inventory:} no hardware configuration $H$ in the available inventory can satisfy the aggregate resource requirements when $S$ is included;
\item \textbf{System incompatibility:} $S$ is incompatible with one or more systems among the fixed choices in $C \setminus C'$ or any feasible replacements within $C'$.
\end{enumerate}

\paragraph{Detailed explanation.}
We now argue that these categories are exhaustive. By construction, any constraint involving a system $S$ in \sysname falls into one of three classes: (i) constraints imposed by the workload specification $W$, (ii) constraints on hardware resources, or (iii) constraints involving other systems.

\textbf{Case 1 (Workload constraints).}
Consider first the case where $S$ has no intrinsic constraints beyond the role it fulfills. In this setting, the feasibility of selecting $S$ depends solely on its compatibility with $W$. Thus, $S$ can be excluded only due to a mismatch with the workload specification, corresponding to conditions (1) or (2).

\textbf{Case 2 (Hardware constraints).}
Next, suppose $S$ only imposes constraints on hardware, defined over the inventory $\mathcal{H}$. A necessary condition for selecting $S$ is the existence of a hardware assignment $H \subseteq \mathcal{H}$ that satisfies both the requirements of $S$ and the aggregate demands of the configuration. Even if such hardware exists in isolation, other choices in $C \setminus C'$ may restrict the feasible hardware set to $H' \subseteq \mathcal{H}$. If $H \cap H' = \emptyset$, then no feasible deployment exists that includes $S$. This corresponds to condition (3).

\textbf{Case 3 (System constraints).}
Finally, consider the case where $S$ imposes constraints on other systems. These constraints may directly prohibit compatibility with systems in $C \setminus C'$ or indirectly restrict feasible replacements within $C'$. In either case, the exclusion of $S$ arises from incompatibility with other system choices, corresponding to condition (4).

\paragraph{Conclusion.}
The above cases are exhaustive by the construction of \sysname’s constraint model. Therefore, any explanation for why a system $S$ is not selected must fall into one or more of the four categories listed above. In practice, multiple conditions may apply simultaneously.

\subsection{Implementation}
Explainability extends \sysname to provide reasoning whenever a set of choices do not match the user's expectation. In order to formalize the user's expectation, we can make use of \textbf{Orderings} that already exist as part of \sysname. The following process describes the steps taken by the tool when the explainability subroutine is invoked. \Circled{1} The subroutine assesses the current output and the user's input query consisting of a role and objective to determine if the relevant ordering has a top priority system that wasn't picked. If it determines there isn't any, the configuration is optimal on the basis of the current constraints and orderings. \Circled{2} If it determines a higher priority system, it adds new temporary constraints that fix all other hardware choices for devices and system choices for roles and replaces the system in question (marked as lower priority) with the higher ranked system in the partial ordering. While this enforces a very strict set of of constraints, we provide the option for users to specify a subset of choices that they wish to make flexible as an user argument. \Circled{3} The solver uses the new set of constraints to attempt to find a satisfiable configuration. In the case where it can find it, we present a side-by-side contrast of the two (the original output without the top-ranked system and the new output with the top-ranked system). While the top ranked system might be chosen for a particular role, by definition of our orderings and picking optimal configurations, it is guaranteed that the new output will have at least one other ordering that will have a suboptimal system chosen (a system that is lower ranked than its counterpart in the original output). \Circled{4} In the case where there is no satisfiable output from the current subroutine, we make use of the UNSAT core, breaking down the unsatisfiable constraints into a human-readable format necessary for interfacing with the user, but still precise enough to point to the exact constraints causing the unsatisfiable result. 

We make an important design choice based on our observations that the UNSAT core only includes very large and complex constraints that are usually conjunctions of multiple constraints encoding diverse properties across the set of choices. This is an attribute of the Z3 solver's internal workings that only includes the top-level constraints in its core. Considering the intended number of constraints and the search space of \sysname, it would not make sense for the user to have only the top-level constraints provided as output. We hence include an extra step in breaking down these constraints into simpler ones through a recursive process: every conjunction, negation, and implication clauses is recursively broken down into atomic clauses that cannot be broken down anymore. We filter these by running our solver over \textbf{only} the UNSAT core, but now consisting of many more, smaller constraints that can be directly mapped to a simple dependency or property of a system or hardware. While this introduces another intervention of the solver, we reason that the input is constrained only to the UNSAT core, and with smaller clauses to evaluate on, the overhead is insignificant (we allow the possibility of limiting the atomicity of the clauses for faster convergence which in turn results in a more complex final output to parse).

\begin{listing}[!t]
\begin{lstlisting}
EXPLAINING
High priority systems for ML_Training than PLB : ['PacketSpray']
...
CHOICES

Solution PacketSpray for workload ML_Training == True
-------
Solution RDMA for workload ML_Training == True
-------
@@@@@@@@@@@@@@@@@@@@@@@@@
CONSTRAINTS
Implies ( Solution RDMA for workload ML_Training , 
        And ( And ( pod1IN-POD_0IN-RACK_0_DEVICE_TYPE.COMPUTE_RDMA , pod1IN-POD_0IN-RACK_1_DEVICE_TYPE.COMPUTE_RDMA , 
        ...
        )))
-------
Implies ( Solution PacketSpray for workload ML_Training , 
        And ( pod1IN-POD_0IN-RACK_0_DEVICE_TYPE.
        COMPUTE_NIC_Reorder_Buffer > 20 , pod1IN-POD_0IN-RACK_1_DEVICE_TYPE.
        COMPUTE_NIC_Reorder_Buffer > 20 , 
        ...
        ))
-------
Implies ( 0 ==
        Hardware ID for pod1IN-POD_0IN-RACK_0_DEVICE_TYPE.COMPUTE , 
        pod1IN-POD_0IN-RACK_0_DEVICE_TYPE.COMPUTE_RDMA ==
        False ) 
-------
Implies ( 1 ==
        Hardware ID for pod1IN-POD_0IN-RACK_0_DEVICE_TYPE.COMPUTE , 
        pod1IN-POD_0IN-RACK_0_DEVICE_TYPE.COMPUTE_RDMA ==
        False ) 
...
\end{lstlisting}
\vspace{0.1in}
\caption{Explainability Output}\label{l:exp-output}
\vspace{-10pt}
\end{listing}

\paragraph{Improving the legibility of the explanation} The raw output of \sysname’s explanation is a filtered version of the UNSAT core, designed to be readable even for users unfamiliar with \sysname’s internals. To support this, \sysname allows experts to associate natural language labels with individual constraints in system and hardware specifications, which are then used to improve the clarity of explanations. Listing~\ref{l:exp-output} shows an example of such raw output for the case study in Section~\ref{sec:case_study_2}. While this output is already interpretable, we find that LLMs can further improve usability by summarizing it into a more concise form. Table~\ref{tab:packetspray-rdma} shows the corresponding LLM-generated summary based on \sysname’s raw explanation.

\section{Additional Case Studies}
\label{app:case_studies}

We demonstrate \sysname's expressiveness by showing a more elaborate set of illustrative case-studies. We follow the same setup as described in Section \ref{s:casestudies}, considering iterative revisions and tradeoffs per design and query step. 

\subsection{Navigating the complex tradeoff space}\label{s:casestudies:tradeoffs}

\begin{listing}[t]
\begin{lstlisting}
inference_app = Workload(name = "Inference"
  deployed_at = racks[0:3],
  properties = [dc_flows, short_flows, high_priority, ml_inference],
  peak_cores = 2800, peak_bandwidth = 30,
  average_cores = 800, average_bandwidth = 10,
  num_flows = 10)
  
inference_app.set_performance_bound(objective = load_balancing, atleast = PacketSpray)
inference_app.set_performance_bound(objective = monitoring, solves = monitoring.detect_queue_length)
inference_app.exempt_role(virtual_switch)

Optimize(inference_app, latency, 1)
Optimize(inference_app, throughput, 2)
Optimize(TOTAL_COST, 3)
Optimize(inference_app, monitoring, 4)
---
Time to generate output: 3.4 seconds
** Systems deployed **
Inference workload -
1. Network stack: Netchannel 
2. Congestion control: BFC
3. Transport protocol: TCP
4. Network load balancer: PacketSpray
5. Network monitoring: Sonata

** Hardware deployed **
Racks[0] -
1. Server: "Intel Xeon 96 cores, 128GB DDR4, CX5 NIC, 4x-NVIDIA V100 16GB",
2. TOR switch: "Tofino1 2Tbps, 2 pipeline",
3. Server-Tor cable: "100Gbps Ethernet"
...
\end{lstlisting}
\vspace{0.1in}
\caption{Description of the ML inference workload with \sysname, and \sysname's output.}\label{l:microservices-workload-app}
\vspace{-10pt}
\end{listing}
\noindent\textbf{Single workload.} We begin by showing how to use \sysname to design a cluster to run a single low-latency ML inference application, consisting of the same roles and hardware as the ML training case in Section \ref{s:casestudies}. The results are given in listing \ref{l:microservices-workload-app}, consisting of nuances encoded as constraints for systems.
For example, the encoding of the packet spraying system captured that with perfect load balancing, larger reorder buffer in the NICs are required. \sysname picked such a NIC according to its library of hardware specifications. Similarly, monitoring queue lengths is an objective only a few systems, e.g., Simon~\cite{simon} (requiring SmartNICs and described in Listing~\ref{l:simon}), Sonata~\cite{sonata}, and Marple~\cite{marple} (latter two requiring programmable switches), satisfy. Thus, \sysname selects a programmable switch (line 29) and deploys Sonata; this is because deploying a programmable switch allowed for the use of BFC~\cite{bfc} (line 21) to improve latency. 

\begin{listing}[t]
\begin{lstlisting}
def simon_constraints(workload):
  devices = workload.get_deployed_devices()
  computes = devices.get(DEVICE_TYPE.COMPUTES)
  constraints = []
  for compute in computes:
    constraints.append(compute.has("NIC_TIMESTAMPS"))
  
  constraints.append(computes.allocate_globally("cores", SIMON_CPU_FACTOR*workload.num_flows))
  return constraints

SIMON = System(id = "SIMON",
    role = MONITOR,
    deployment_constraints = simon_constraints,
    solves = [
        monitoring.capture_delays,
        monitoring.detect_queue_length
    ]
)
SIMON.add_warning(lambda workload: workload.get_deployed_links().network_load > 40, "Simon does not have benchmarks beyond 40Gbps link speeds and it is being deployed for that case")

Ordering(objective = monitoring, SIMON, better_than = PINGMESH)
Ordering(objective = ease_of_deployment, PINGMESH, better_than = SIMON)
\end{lstlisting}
\vspace{0.1in}
\caption{System description of SIMON.}\label{l:simon}
\vspace{-10pt}
\end{listing}

\noindent\textbf{Adding ML training.} 
We consider the evolution of the workload by considering how the deployment might change if the same cluster also had to support an ML training workload in addition to the ML inference one described above.
After encoding the ML training workload, the architect specifies that the deployment should include virtualization between the two workloads. This triggers several changes: 
first, \sysname notices that the CPU load has increased to require $3,200$ cores, whereas the previous design required $2,800$. 
However, the architect specifies that the deployment site has limited rack space, so it's not possible to deploy more copies of the same server.
Thus, \sysname picks larger 128-core servers, which can accommodate not only the new workload, but also the one extra core per server that Snap burns for scheduling.
Note that since the workloads and systems in this example specify their CPU load numerically, \sysname can reason about a constraint (following common practice) to keep peak CPU utilization below 80\% of the total cores.
Second, \sysname deploys a network virtualization system, Andromeda~\cite{andromeda}. \sysname picks Andromeda in this case because the server upgrade freed up enough cores to support it, while other virtualization systems require SmartNICs~\cite{vfp, nitro}.
Andromeda then includes a requirement that applications use the Snap~\cite{snap} networking stack instead of NetChannel.
In cases such as this one, \sysname provides value because it tracks nuances across a high-dimensional optimization space.

\paragrapha{Controlling Cost}
The previous design used programmable switches for congestion control (BFC~\cite{bfc}) and monitoring (Sonata~\cite{sonata}). 
Now, the architect considers a deployment that considers the high cost of expensive programmable switches, and removes these from the input inventory.
Now, \sysname uses Swift~\cite{swift} for congestion control, and switches to Simon~\cite{simon} for monitoring. 
Note that Simon, as an end-host-based system, offers lower fidelity monitoring than Sonata, but \sysname picks it anyway because the architect specified cost as a higher-priority objective than monitoring.
Also note that PingMesh~\cite{pingmesh}, though it is even lower cost, is not suitable here because it does not satisfy the \texttt{detect\_queue\_length} objective. \sysname decides to  use  the SWIFT~\cite{swift} congestion control algorithm since it doesn't require an expensive programmable switch's support. For monitoring it notices that it already switched to FPGA-based smart NICs, and therefore picks Simon~\cite{simon} which is CPU efficient. We finally see that with increased network load, PacketSpray cannot be selected anymore, as the latency objective requires a larger set of reorder buffer for the NICs.

\subsection{Practical and incremental deployments}
Section \ref{s:casestudies:tradeoffs} showed the complexity of the tradeoffs between systems, objectives, and the hardware, and the nuances involved in choosing between them. 
It also demonstrates the subtle yet practical impact of  changes in the workload and the desired scale of the network and how it changes the solution space and the selected solutions. In practice, it is not possible for a cloud provider to keep changing the deployments so rapidly -- it is time-consuming, disruptive, and costs money. 
Instead, most operators have an extensible design, but use the same hardware and systems with slight modifications to satisfy the new requirements. We designed \sysname to allow architects to express this practicality, and showcase this with the following three examples.

\noindent\textbf{Unwilling to upgrade devices.}
The architect can specify which devices they are unwilling to change and which systems they want to retain in a deployment plan. In the previous example, if the architect was unwilling to change the servers that they had already purchased while deploying the ML inference app, they could encode this constraint as:
\begin{lstlisting}
for compute in racks[0].get(DEVICE_TYPE.COMPUTE):
    compute.assign_hardware("Intel Xeon 96 cores, CX5 NIC")
\end{lstlisting}
\noindent
With this constraint, \sysname does not use Snap since the deployment is constrained on CPU cores. Instead, it chooses Demikernel's RDMA stack, which is possible since the deployment already includes RDMA-capable Mellanox CX-5 NICs.
Further, because \sysname picked RDMA, it also deploys DCQCN instead of BFC, and deploys PFC because it is a requirement for DCQCN. \sysname also changes the switch's hardware to a fixed-function switch, but one with enough physical queues to support isolation between TCP and RDMA traffic.

\noindent\textbf{Avoiding special hardware.}
Consider a particular case when an architect is unwilling to use non-standard hardware such as programmable switches or hardware offloads.
To express this, the architect has two options.
First, they can change orderings on the \texttt{ease\_of\_deployment} objective, indicating that it's harder for them to use systems that require fancier hardware. Then, they can optimize for ease of deployment with higher priority than other orderings.
Second, just as the architect can specify hardware to continue using, they can specify hardware they are unwilling to deploy.

\subsection{Adopting new hardware}
To consider how \sysname can model an evolving deployment, we consider the evolution of SmartNICs and their use in deployments over time. 
We show that using \sysname, architects can reason about when it makes sense for them to change their hardware deployments.

The first widely adopted cloud SmartNICs were based on Intel FPGA platforms. They enabled offloading virtual networking features to save CPU cores, but required expertise to operate and deploy, and were expensive.
Before the introduction of these SmartNICs, a \sysname NIC schema might only encode a few properties such as the supported line rate.
Then, to model these SmartNICs in \sysname, architects would update server schemas to reflect the availability of the FPGA-based virtualization offload.
Then, systems experts could define systems that made use of these FPGA-based SmartNICs such as VFP~\cite{vfp} in 2017.

Later, in 2019 Amazon published Nitro~\cite{nitro}, a SmartNIC based on ARM cores rather than FPGAs. These SmartNICs could offload not only virtual networking but also some transport offloads. \sysname would express this difference using orderings: the Nitro system would not only satisfy the virtual networking role, but also be ordered preferentially to other systems without transport offloads.

Thus, by encoding the features of systems and hardware as trends evolve, \sysname can help architects reason about the tradeoffs involved in their deployment choices.
Naturally, this must depend on the architect's requirements, expertise, and preferences; one operator might view deploying FGPA-based SmartNICs as having a prohibitively low ease-of-deployment, while another more used to working with FPGAs might not have this issue.
\sysname thus allows architects to express these instance-specific preferences to reason about their deployments while also benefiting from community-sourced specifications of other systems, hardware, and workloads.

\end{document}